\documentclass[12pt,a4paper]{article}
\begin{document}
  
\title{Technique for measuring the parameters of polarization of an ultrasonic wave}

\author{A. M. Burkhanov, K. B. Vlasov, V. V. Gudkov, and B. V. Tarasov}

\maketitle

Among physical phenomena consisting in variation of the polarization of a shear ulrasonic wave the acoustic analogs of the Faraday and the Cotton-Mouton effects are  investigated at present (see \cite{kitt} and \cite{vl_59} -- the first theorectical papers, \cite{morse-gav_59} -- discovery of rotation of the polarization, and \cite{matth} --\cite{ta-bur-vl_96} --  some    experiments). They are observed when initially linearly polarized ultrasonic wave  propagates inside  a bulk specimen and are due to interaction between elastic and  magnetic subsystems or conduction electrons. Quantitative characteristics of the effects are  polarization parameters: $\varepsilon$ -- the ellipticity which modulus is the ratio of the minor and major ellipse axes and $\phi$ -- the angle of rotation of the polarization plane  or, more correctly,  of the major ellipse axis if  $\varepsilon\neq0$. Most of recent experiments on polarization phenomena were performed with the use of phase-amplitude methods. A review of them is given in Ref.~\cite{gu-tar_98}. 

Besides, a  phenomenon is considered as the acoustic analog of magneto-optic Kerr effect if variation of the polarization occurs while reflection of the wave from an interface between magnetic medium and isotropic non-magnetic one. It  was predicted by Vlasov and Kuleev \cite{vl_ku-68} in 1968, however, there was no papers  yet about experiments in which both the parameters characterizing the polarization, $\varepsilon$  and $\phi$, were measured. 

We have completed such an experiment and the results will be published soon. While performing it  we found that  a very small variations of a high level signal took place and  came to a conclusion  that amplitude variant of a technique should be more suitable here. 

First amplitude technique for a precise measurement of  $\phi$ was introduced by Boyd and Gavenda \cite{boyd-gav_66}. Its aplicability  was limited to the case where  $\varepsilon \approx0$.  Though, we developed  an amplitude method  free of this restriction for measuring $\phi$    as well as $\varepsilon$. A description of the technique is the subject of this paper.

The method  consists of measuring the amplitude of the voltage,  $V(H)$, on the receiving transducer  at a certain $B_1$ relative to an initial $B=B_0$ using three different angles for the receiving transducer $\psi$  with futher processing of the data with the formulas (\ref{8}), (\ref{tech1.9}), and (\ref{tech1.10}) presented  below. It can be used for investigating  the acoustic analogs of the Faraday  and the Cotton-Mouton effects as well.

\vspace*{1cm}
\label{A1}
 
 A periodic motion of the volume element over an elliptic trajectory can be repersented  with the help of amplitudes, $u^\pm$, and  phases, $\varphi^\pm$,  of circular  elastic vibrations. Introducing a parameter
\begin{equation}
p= (u^-/u^+) \, e^{i(\varphi^- -\varphi^+)},
  \label{7}
\end{equation}
expressions for $\varepsilon$ and $\phi$ have the form:
\begin{eqnarray}
\varepsilon  =  \frac{1 - |p|}{1 + |p|}, \ \ \ \ \
\phi  =  -\,\frac{1}{2}\, {\rm Im}\, \left[\ln(p)\right].  \label{8}
\end{eqnarray}

Projection of the   elastic vibrations to polarization direction of the receiving transducer can be written as  follows: 
\begin{eqnarray}
u_r (t)={\bf u} \cdot {\bf e} _r = 
{\rm Re} \left\{u^+ \exp\left[i(\omega t - \varphi^+ -\psi)\right]\hspace*{1cm}
 \right. \nonumber\\ \left. + u^- \exp\left[-i(\omega t- \varphi^- + \psi)\right] \right\}, 
\label{9}
\end{eqnarray}
where * designates the complex conjugate,  ${\mathbf e}_r$ is  unit vector of the direction of the polarization of the receiving transducer, $\psi$ is the angle between this direction and the plane of incidence, $\omega$ is frequency, and $t$ is time. $u_r$  excite an ac voltage $V\!\cos(\omega t- \alpha) = \eta u_r$  (where $\eta ^2$ is the coefficient of transformation of elastic vibration energy into electric field energy, and $\alpha$\ is a phase constant). Using Eq.\  (\ref{9}) we have
\begin{eqnarray}
\lefteqn{\frac{V}{\eta}\left[\cos\omega t\cos\alpha +\sin\omega t\sin\alpha\right]}  \hspace{0.5cm} \nonumber \\ &\!\!\! = &\!\!\!\, \left[u^+ \cos(\varphi^+ +\psi) + u^- \cos(\varphi^- - \psi)\right]\cos\omega t \nonumber \\
 && + \left[u^+ \sin(\varphi^+ +\psi)+ u^- \sin(\varphi^- - \psi)\right]\sin\omega  t. \label{11}
\end{eqnarray}

Since Eq.\ (\ref{11}) is valid for arbitrary $t$, it may be transformed into two equations:
\begin{eqnarray}
\frac{V}{\eta} \cos\alpha \,\,&\!\!\! = &\!\!\!  \,u^+ \cos(\varphi^+ +\psi)
+ u^- \cos(\varphi^- - \psi),   \label{12}\\
\frac{V}{\eta} \sin\alpha \,\,&\!\!\! = &\!\!\! \,u^+ \sin(\varphi^+ +\psi)+ u^- \sin(\varphi^- - \psi).  \label{13} 
\end{eqnarray}

Multiplying Eq.\ (\ref{13}) by $i$ and adding the result to Eq.\ (\ref{12}) we obtain 
\begin{eqnarray}
\frac{V}{\eta} 
e^{ i\alpha}
=\, u^+ 
e^{i(\varphi^+ +\psi)}
+ u^- 
e^{i(\varphi^- - \psi)}. \label{14}
\end{eqnarray}

The method suggested here for determining the  polarization of the reflected wave consists of measuring the amplitude of the signal  at a certain $B_1$ relative to an initial $B=B_0$ using three different angles for the receiving transducer: $\psi_1, \psi_2$, and $\psi_3$. We assume that $\varepsilon (B_0)=0$ and $\phi (B_0)=0$.  Relevant equations for the two different values of $B$ and three of $\psi$ may be obtained by making the appropriate substitutions into Eq.\  (\ref{14}). Introducing indexes  $j=0, 1$ for the two values of $B$ and $k=1,2,3$ for the three values of $\psi$ for $V_{kj}$, $\alpha_{kj}$, $u^\pm_j$, and $\varphi^\pm_j$ we have:
\begin{eqnarray}
\frac{V_{10}}{\eta} \, e^{i\alpha_{10}}
=\,u^+_0
e^{i\left(\varphi^+_0+\psi_1\right) }
+ u^-_0
e^{i\left(\varphi^-_0- \psi_1\right)}, \label{30} \\
\frac{V_{11}}{\eta} \, e^{i\alpha_{11}}
=\,u^+_1
e^{i\left(\varphi^+_1+\psi_1\right)} + u^-_1
e^{i\left(\varphi^-_1- \psi_1\right)} \label{31} . 
\end{eqnarray}
Dividing  Eq.\ (\ref{31})  by (\ref{30}), we obtain
\begin{eqnarray}
\frac{V_{11}}{V_{10}} e^{i\left(\alpha_{11} - \alpha_{10}\right)} 
\,= F^+_1 e^{i \psi_1} + F^-_1 e^{-i \psi_1},   \label{33} 
\end{eqnarray}
where
\begin{eqnarray}
F^\pm_1
\equiv \frac{u^\pm_1 \, e^{i\varphi^\pm_1}}{u^+_0\exp [i(\varphi^+_0 +\psi_1) ]+ u^-_0 \exp [i(\varphi^-_0 -\psi_1)]} .   \label{39}
\end{eqnarray}

Similar equations for $\psi =\psi_2$ have the form
\begin{eqnarray}
\frac{V_{21}}{V_{10}} 
e^{i\left(\alpha_{21} - \alpha_{10}\right)}\delta_2\, e^{i\lambda_2} 
= F^+_1\, e^{i\psi_2} + F^-_1\, e^{-i \psi_2}, \label{35}
\end{eqnarray}
where $\lambda_2$ and $\delta_2$ describe variations in phase and amplitude of the signal, respectively, caused by differences in transducer coupling to the sample while changing $\psi$ from $\psi_1$ to $\psi_2$.

One more change in $\psi$ gives the following equations in addition to (\ref{33}) and (\ref{35}):
\begin{eqnarray}
\frac{V_{31}}{V_{10}} 
e^{i\left(\alpha_{31} - \alpha_{10}\right) } \,\delta_3\, e^{i\lambda_3}
= F^+_1\, e^{i \psi_3} + F^-_1\, e^{-i \psi_3} .   \label{37} 
\end{eqnarray}
Here $\delta_3$\ and $\lambda_3$\ have the same origin as $\delta_2$\ and $\lambda_2$, but correspond to changing $\psi$ from $\psi_1$ to $\psi_3$.

After multiplying the left and right sides of Eqs.\ (\ref{33}), (\ref{35}), and (\ref{37}) by their complex conjugates we obtain
\begin{eqnarray}
\left(\frac{V_{11}}{V_{10}}\right)^2
= \left| F^+_1\right|^2+ \left| F^-_1\right|^2 + 2\left| F^+_1\right| \left| F^-_1\right|\cos \left(\Delta \varphi_1 + 2\psi_1\right), 
\label{tech1.1} \\
{\left(\frac{V_{21}\, \delta{_2}}{V_{10}}\right)^2}
 =\left| F^+_1\right|^2+ \left| F^-_1\right|^2 + 2\left| F^+_1\right| \left| F^-_1\right|\cos \left(\Delta \varphi_1 + 2\psi_2\right), 
\label{tech1.2} \\
{\left(\frac{V_{31}\, \delta_{3}}{V_{10}}\right)^2}
= \left| F^+_1\right|^2+ \left| F^-_1\right|^2 + 2\left| F^+_1\right| \left| F^-_1\right|\cos \left(\Delta \varphi_1 + 2\psi_3\right), 
 \label{tech1.3}
\end{eqnarray}
where  
\begin{eqnarray}
\Delta \varphi_1=\varphi^+(B_1) - \varphi^-(B_1), \label{tech1.4}
\end{eqnarray}
and, due to the assumption of $\varepsilon (B_0)=0$ and $\phi (B_0)=0$,
\begin{eqnarray}
\delta_i = \frac{V_{10}\cos \left(\psi_i\right)}{V_{i0}\cos \left(\psi_1\right)}. \label{at5}
\end{eqnarray}

These operations are necessary to remove the phase $\alpha_{kj}$ from our equations since amplitude  is the only parameter measured  in this variant of a technique. We divide both sides of Eqs.\ (\ref{tech1.1})--(\ref{tech1.3}) by $\left|F^+_1\right|\left| F^-_1\right|$ to obtain
\begin{eqnarray}
\left| p_1\right|^{-1}+\left| p_1\right| +2\cos\left[2(\phi_1-\psi_1)\right]&\!\!\! = &\!\!\!\frac{\left(V_{11}/V_{10}\right)^2}{\left| F^+_1\right| \left| F^-_1\right|},        \label{tech1.6} \\
\left| p_1\right|^{-1}+\left| p_1\right| +2\cos\left[2(\phi_1-\psi_2)\right]&\!\!\! = &\!\!\!\frac{\left(V_{21}\, \delta_2/V_{10}\right)^2}{\left| F^+_1\right| \left| F^-_1\right|},  \label{tech1.7} \\
\left| p_1\right|^{-1}+\left| p_1\right|+2\cos\left[2(\phi_1-\psi_3)\right] &\!\!\! = &\!\!\! \frac{\left(V_{31}\, \delta_3/V_{10}\right)^2}{\left| F^+_1\right| \left| F^-_1\right|}, \label{tech1.8}
\end{eqnarray}
where $p_1\equiv p(B_1)$ . 

Thus we have three equations with three unknowns, namely $\left| F^+_1\right| \left| F^-_1\right|$, $\left| p_1\right|$, and $\phi_1$. The latter two are the parameters we are interested in and corresponding solutions of the system have the form
\begin{eqnarray}
\phi_1 \,&\!\!\! = &\!\!\!\, \frac{1}{2}\tan^{-1} \Bigl\{\bigl[\left(V_{21}^2\, \delta_2^2 - V_{31}^2\, \delta_3^2\right) \cos 2\psi_1  \nonumber \\
& & \mbox{\hspace{.3cm}} + \left(V_{31}^2\, \delta_3^2 - V_{11}^2\right) \cos 2\psi_2 + \left(V_{11}^2 - V_{21}^2\, \delta_2^2\right) \cos 2\psi_3\bigr] \nonumber \\ 
& & \times\bigl[\left(V_{21}^2\, \delta_2^2 - V_{31}^2\, \delta_3^2 \right) \sin 2\psi_1 + \left(V_{31}^2\, \delta_3^2 - V_{11}^2\right) \sin 2\psi_2 \nonumber \\
& & \mbox{\hspace{.3cm}} + \left(V_{11}^2 - V_{21}^2\, \delta_2^2\right) \sin 2\psi_3 \bigr]^{-1}\Bigr\} \label{tech1.9}
\end{eqnarray}
and
\begin{eqnarray}
\left| p_1\right|\,=\, \frac{a_1}{c_1} \pm \left[\left(\frac{a_1 }{c_1}\right)^2-1\right]^{1/2},    \label{tech1.10}
\end{eqnarray}
where
\begin{eqnarray}
a_1 \,&\!\!\! = &\!\!\! \, V_{11}^2 \sin\left[2(\psi_2-\psi_3)\right] + V_{21}^2\, \delta_2^2 \sin\left[(2(\psi_3-\psi_1)\right] \nonumber \\
& & \mbox{\hspace{.7cm}} + V_{31}^2\, \delta_3^2 \sin\left[2(\psi_1-\psi_2)\right] \cos 2\phi_1, \nonumber \\
c_1 \,&\!\!\! = &\!\!\!\, \left(V_{21}^2\, \delta_2^2 - V_{31}^2\, \delta_3^2\right) \sin 2\psi_1 + \left(V_{31}^2\, \delta_3^2 - V_{11}^2\right) \sin 2\psi_2  \nonumber \\
& & \mbox{\hspace{.7cm}} + \left(V_{11}^2 - V_{21}^2\, \delta_2^2\right) \sin 2\psi_3. \nonumber 
\end{eqnarray}
The $(-)$ sign should be taken before the square root in Eq.\,(\ref{tech1.10}), since it alone allows $\left| p_1\right| =0$ and therefore $\varepsilon =1$.

\end{document}